\documentclass[11pt]{article}
\usepackage{moriond,epsfig}

\def\be{\begin{equation}}
\def\ee{\end{equation}}
\def\bea{\begin{eqnarray}}
\def\eea{\end{eqnarray}}

\begin{document}
\vspace*{4cm}
\title{Production of Isolated Photons in Deep Inelastic Scattering}

\author{A.~Gehrmann-De Ridder}

\address{Institute for Theoretical Physics, ETH, CH-8093 Z\"urich, Switzerland}

\maketitle
\vskip -5 cm
\centerline{\includegraphics[width = 3.7cm]{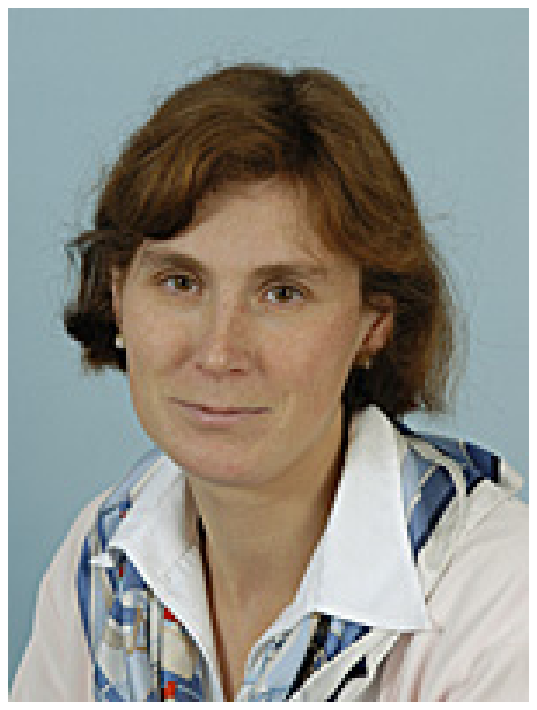}}

\abstracts
{We present here the predictions obtained from a 
calculation of the inclusive isolated photon production cross section in deep
inelastic scattering.
The results are compared with 
the cross section measurement of the ZEUS collaboration and found 
in good agreement with all aspects of it.
Furthermore, a way of measuring the quark-to-photon fragmention 
function in DIS is also briefly outlined.}


\section{Introduction}
The ZEUS collaboration at DESY HERA reported recently a 
measurement~\cite{zeus} of 
the inclusive production cross section for isolated photons in deep 
inelastic scattering (DIS).
The normalization of the cross section obtained from the event generators 
HERWIG~\cite{herwig} and PYTHIA~\cite{pythia} was too high by factors 7.9 and 
2.3 respectively. Even after normalizing the total event rate, 
none of these programs was able to describe all kinematical 
dependencies of the measured cross section.
A further attempt to describe the ZEUS measurement 
using the photon distribution in the proton was also considered 
in a subsequent study~\cite{mrst}. However, it failed to describe 
the kinematical distribution $\eta_{\gamma}$ correctly \cite{saxon}. 
To investigate the origin of these discrepancies, 
we performed a new calculation \cite{letter} 
of the inclusive isolated photon cross section in DIS.
In the following, we describe the parton level calculation 
and present the resulting predictions choosing the same kinematical 
constraints as in the experimental analysis.
Finally we shall briefly mention the possibility of measuring 
the quark-to-photon fragmentation function in DIS
events.     

\section{The inclusive isolated photon cross section}
In the following, we outline major aspects 
of the parton-level calculation pointing out the presence of  
three different contributions depending whether the photon is radiated from 
a quark or a lepton. We will briefly discuss each contribution separately and 
in particular mention that, when the photon is radiated off a quark,
both hard photon radiation and quark-to-photon fragmentation contributions 
have to be taken into account.

\subsection{The parton-level calculation}

The leading order parton-level process corresponding to the inclusive 
production of an isolated photon in deep inelastic scattering is given by:
$$ q(p_1) + l(p_2) \to \gamma(p_3) + l(p_4) + q(p_5)\;,$$ 
where $q$ represents a quark or anti-quark, and $l$ a lepton or 
anti-lepton. The measurable cross section for lepton-proton 
scattering $\sigma (ep \to e\gamma X)$ 
is obtained by convoluting the parton-level lepton-quark cross 
section $\hat\sigma (eq \to e\gamma q)$ 
with the quark distribution functions in the proton. 
In the scattering amplitudes for this process, the lepton-quark interaction 
is mediated by a virtual photon, and the final state photon 
can be emitted off the lepton or the quark. Consequently, one finds three
contributions to the cross section, coming from the squared amplitudes 
for radiation off the quark ($QQ$) or the lepton ($LL$), as well as the
interference of these amplitudes ($QL$). 
The $QL$ contribution is odd under charge exchange, such that it contributes
with opposite sign to the cross sections with $l=e^-$ and $l=e^+$. 

In the $QQ$ contribution, the photon can have different origins: 
the direct radiation off the quark and the fragmentation of a hadronic 
jet into a photon carrying a large fraction of the jet energy.
When the radiation takes place at an early stage of the hadronization 
the quark and the photon are usually well separated  
from each other. 
However, when the photon is radiated somewhat later during the
hadronization process, it can be emitted collinearly to the primary
quarks giving rise to a collinear singularity. Both processes have 
to be considered. As physical cross sections are necessarily finite, 
this collinear singularity gets factorized into the fragmentation function 
defined at some factorization scale $\mu_{F,\gamma}$, as usual~\cite{AP}.
In the present calculation, we use the phase space slicing method~\cite{gg}
to handle the collinear quark-photon singularity, as described in 
detail in~\cite{kramer,andrew}. 

The fragmentation contribution, besides absorbing the final state collinear 
singularity,  describes non-perturbative physics effects. It is 
characterized by the process-independent quark-to-photon fragmentation
function~\cite{koller} $D_{q\to \gamma}(z,\mu_{F,\gamma})$,
where $z$ is the momentum fraction carried by the photon.
This fragmentation function can not be calculated within perturbative 
methods. Instead it has to be measured experimentally.
This function has, up to now, been directly 
determined only by the ALEPH collaboration at LEP \cite{aleph}. 
This determination was based on a comparison between 
the measured and the calculated \cite{andrew}  $e^+e^- \to \gamma + 1$~ jet 
rate at LEP.   
Other parametrizations of photon fragmentation functions were proposed 
in the literature~\cite{grv,bfg}, in which a resummation of logarithms of 
 $\mu_{F,\gamma}$ is performed. 
In the calculation \cite{letter}, for the
quark-to-photon fragmentation function,  
we use the ALEPH leading order 
parametrization~\cite{aleph} as default which does not incorporates any 
resummation of logarithms, and the BFG (type I)  
parametrization~\cite{bfg}, evaluated for $\mu_{F,\gamma}^2=Q^2$ 
for comparison. 

The factorization scale $\mu_F^2$ for the parton distributions is 
$Q^2=-(p_4-p_2)^2$ for the $QQ$ subprocess and kinematical constraints
 on $Q^2$ ensure the deep inelastic nature of the $QQ$ process.    
In the $LL$ subprocess, where the final state photon is radiated off 
the lepton, the situation is more complicated. 
It is described in detail in \cite{letter}. 
For the $LL$
subprocess we choose $\mu_F$ = max($\mu_{F,{\rm min}}$,$Q_{LL}$), and for the 
$QL$ interference subprocess $\mu_F$ = 
max($\mu_{F,{\rm min}}$,$(Q_{LL}+Q_{QQ})/2$)
\cite{letter}, with,$Q_{LL}^2=-(p_5-p_1)^2$ and $\mu_{F,{\rm min}}=1$~GeV.
To ensure the deep inelastic nature of the LL subprocess,   
the requirement of observing hadronic tracks is of crucial relevance
as it implies that the proton disintegrates.
As explained in this letter \cite{letter}, 
the track requirement translated into 
parton-level variables corresponds to impose a minimal cut on the quark 
rapidity, $\eta_q <3$, which we impose in our calculation.

\subsection{The results}

The isolated photon cross section in deep inelastic scattering 
is defined by imposing a number of kinematical cuts on the final state 
particles. In the ZEUS analysis, 
which combined three data samples \cite{zeus}, these were chosen as follows: 
virtuality of the process, as determined from the outgoing electron 
$Q^2 = -(p_4-p_2)^2 > 35$~GeV$^2$, 
outgoing electron energy $E_e > 10$~GeV and angle 
$139.8^\circ < \theta_e < 171.8^\circ$, outgoing photon transverse energy
5~GeV$<E_{T,\gamma}<$10~GeV and rapidity $-0.7<\eta_\gamma<0.9$. 

The photon, to be called isolated, is required to carry at 
least 90\% of the energy found in a cone of radius $R=1.0$ in the $\eta-\phi$ 
plane around the photon direction. This cone-based isolation procedure 
is commonly used to define isolated photons produced in a hadronic 
environment. A minimal amount of hadronic activity inside the cone 
has to be allowed in order to ensure infrared finiteness of the observable.


For the numerical evaluation of the cross section, we 
use the CTEQ6L leading order parametrization~\cite{cteq} 
of parton distributions. Using the ZEUS cuts and the ZEUS composition of the 
data sample at different energies and with electrons and positrons \cite{zeus}, we obtain a theoretical prediction 
for the isolated photon cross section in DIS
of 5.39~pb, to be compared to the 
experimental value of 5.64$\pm$0.58(stat.)${+0.47 \atop -0.72}$(syst.)~pb.
The total cross section is therefore well reproduced by our 
calculation. 
\begin{figure}[tbh]
 \begin{center}
\epsfig{file=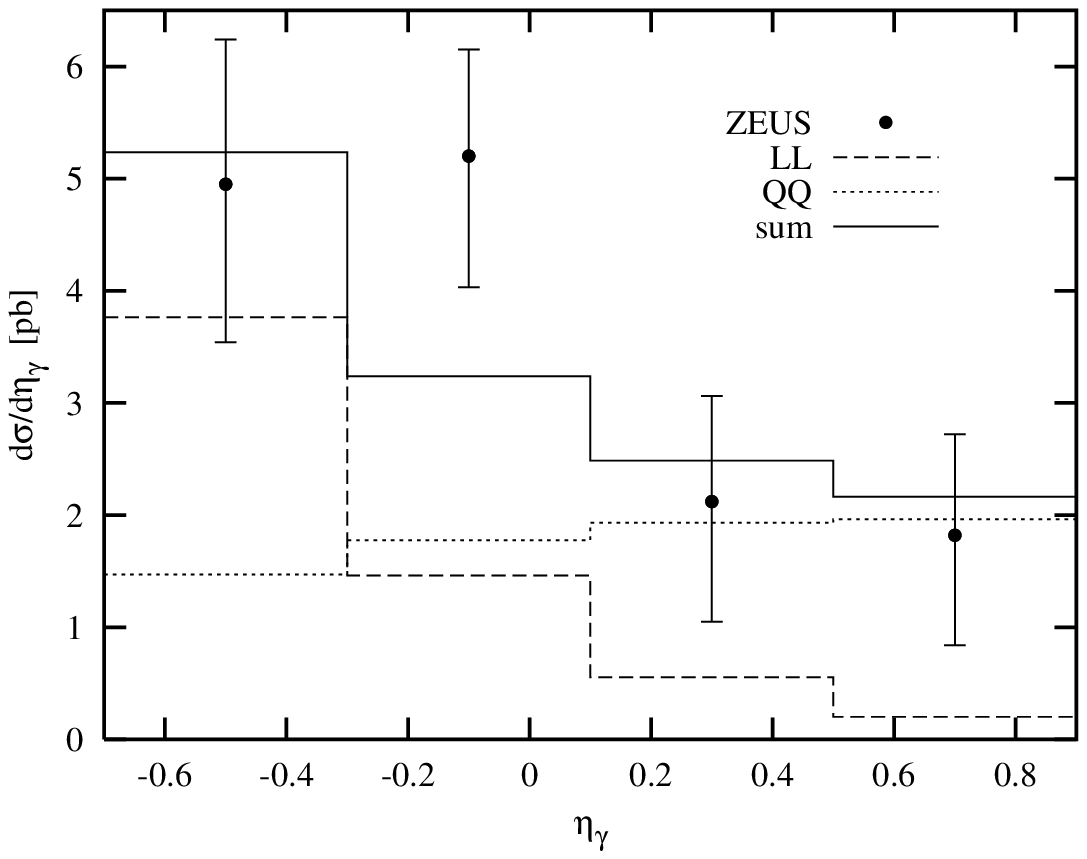,angle=-90,width=6.1cm} 
\epsfig{file=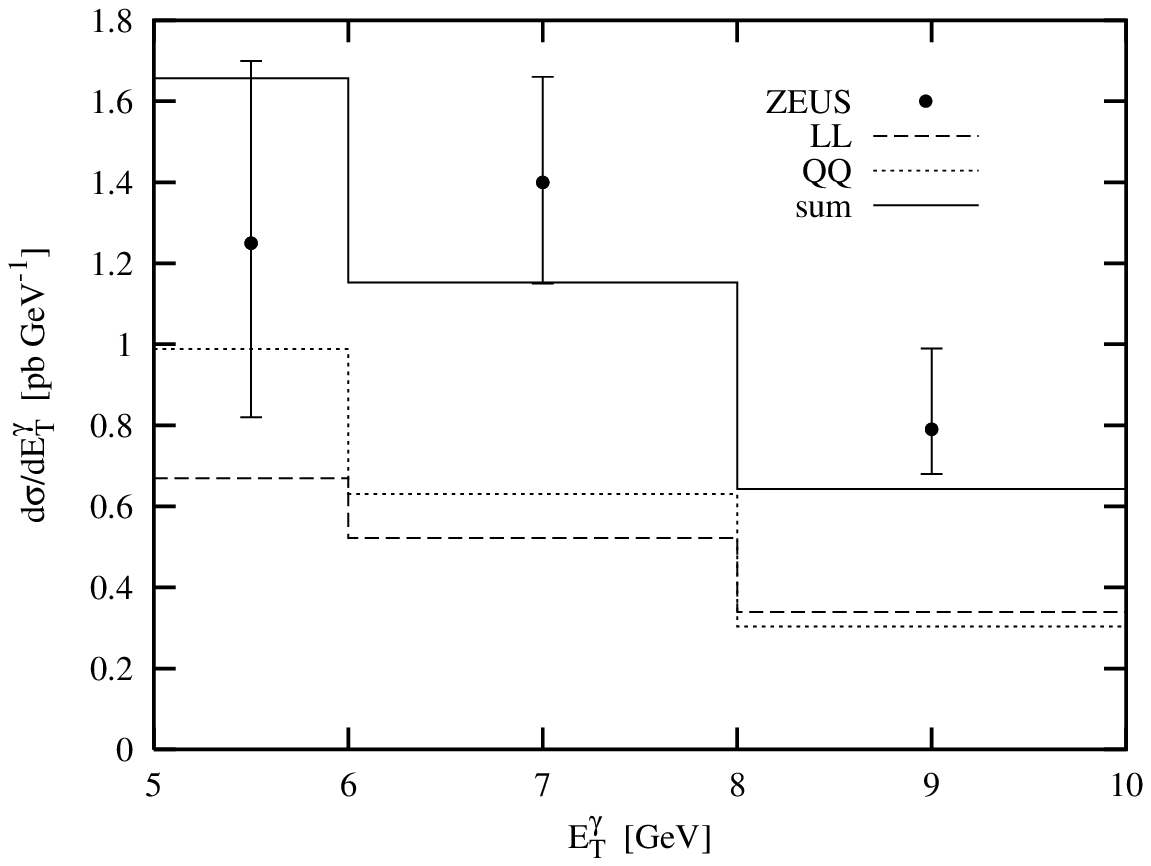,angle=-90,width=6.1cm} 
 \end{center}
\caption{(a) Rapidity distribution and (b) 
transverse momentum distribution of isolated photons, compared 
to ZEUS data.}
\label{fig1}
\end{figure}


Using our leading order calculation, we obtain differential cross sections 
in $\eta_\gamma$, $E_{T,\gamma}$ which are shown in
 Figure \ref{fig1}. Both distributions are found in good agreement 
with the ZEUS data.  


\section{Conclusions and outlook}
In this talk, 
we reported results concerning the production of isolated photons 
in deep inelastic scattering presented in \cite{letter}. 
In this letter, we compare our predictions for the $\sigma(lp \to l\gamma X)$ 
with the ZEUS measurement of this cross section \cite{zeus}.
We found that photon radiation off quarks and leptons contribute 
about equal amounts to this observable.
Since the photon isolation criterion 
admits some amount of hadronic activity around the photon direction, 
small angle radiation off quarks and non-perturbative quark-to-photon 
fragmentation function contributions need to be considered.
Both these effects (large-angle radiation and photon fragmentation) 
are included in our fixed-order parton model calculation, which yields
good agreement with the ZEUS data both in normalization and in shape.

\begin{figure}[tbh]
\begin{center}
\epsfig{file=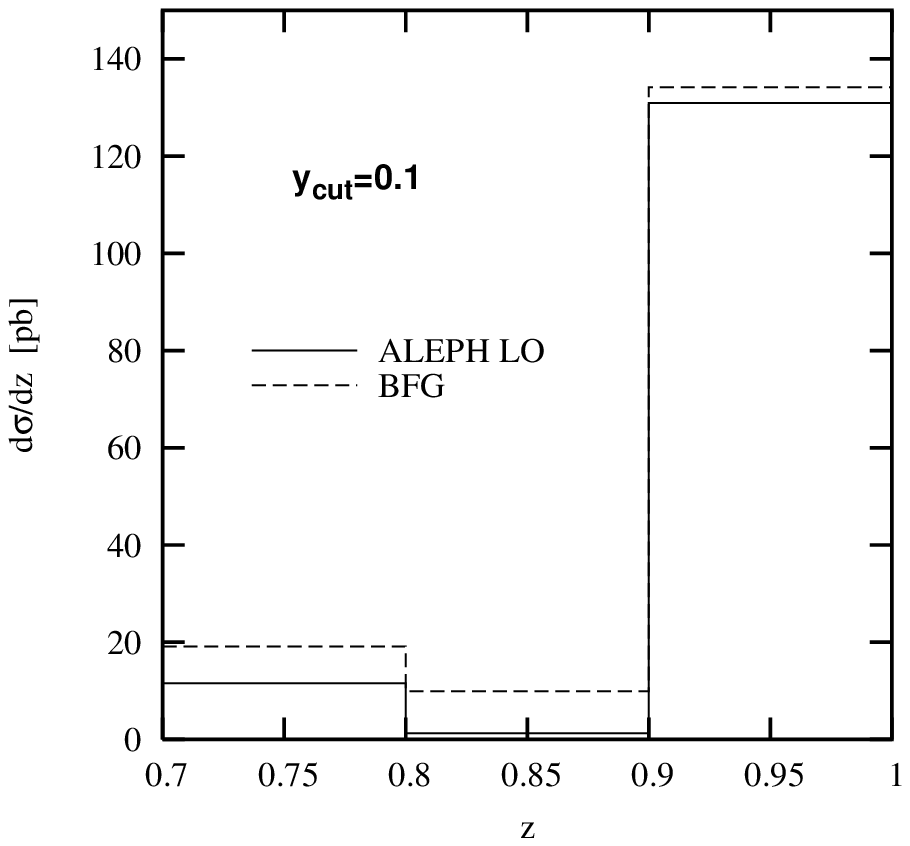,angle=-90,width=5.1cm} 
\epsfig{file=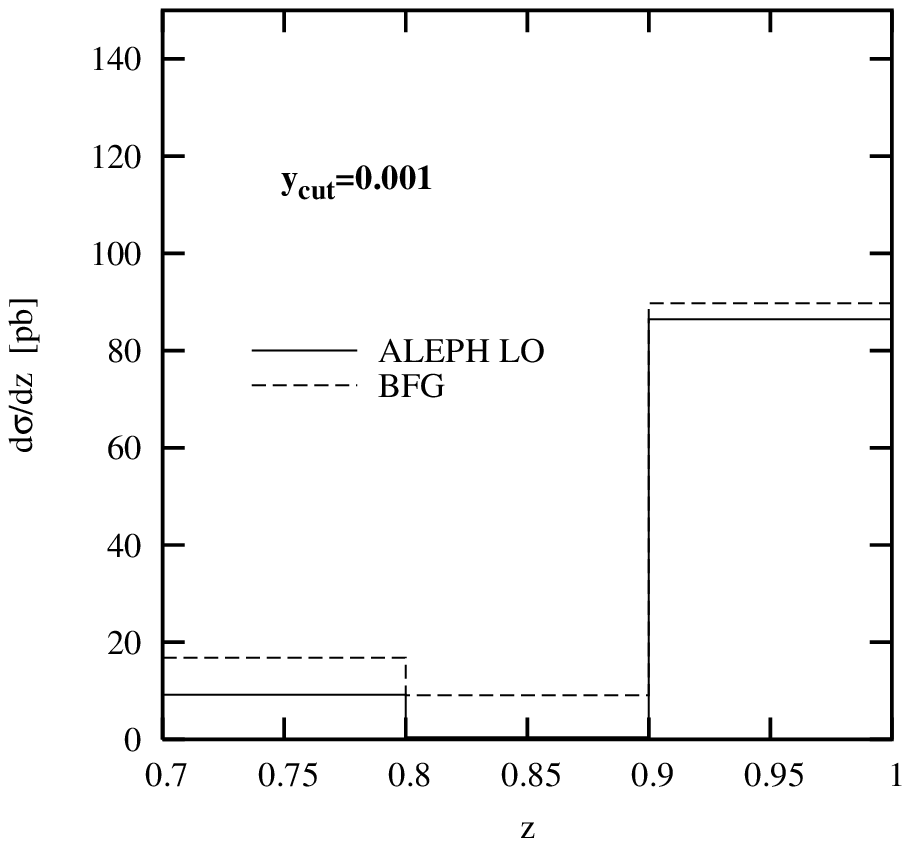,angle=-90,width=5.1cm} 
\end{center}
\caption{ Differential distribution in $z$ $(0.7<z<1)$ 
of the $\sigma(\gamma +(0+1)$ jet) using the exclusive $k_{T}$ algorithm 
with (a) $y_{cut}=0.1$ and  (b) $y_{cut}=0.001$  
for the ALEPH (plain) and BFG (dashed) 
parametrizations of the quark-to-photon fragmentation function} 
\label{fig2}
\end{figure}

By further analyzing the hadronic final state in isolated photon 
production in DIS, one can define more exclusive observables 
like photon-plus-jet cross sections. 
In \cite{paper}, we present the predictions 
obtained from calculating the $\gamma +(0+1)$-jet cross section 
differential in $z$. 
These predictions are obtained using the kinematical cuts appropriate for 
an H1 measurement specified in \cite{carsten} 
and defined using the $k_{T}$  exclusive jet algorithm \cite{k_T}
 in the laboratory frame. For this observable, 
besides the photon jet ($\gamma$) and the proton remnant ($+1$), 
no ($0$) further hadronic jet activity is present in the final state.
Figure~\ref{fig2} displays the theoretical results as bin-integrated 
cross sections for three bins for two different values of the jet resolution
parameter $y_{cut}$.

For a given $y_{cut}$, the predictions for the $\gamma+(0+1)$-jet 
cross section obtained using different parametrisztions shown 
in Figure \ref{fig2} \cite{paper} 
differ considerably; in particular for $0.7<z<0.9$. 
It appears therefore that this observable 
is highly sensitive on the photon FF and would be an appropriate 
observable to measure in view of extracting
the quark-to-photon fragmentation function in DIS.

\section*{Acknowledgments}

This work was supported by the Swiss National Science Foundation 
(SNF) under contract PMPD2-106101.

\end{document}